# Coarsened Lattice Model for Random Granular Systems

R. Piasecki[1])

*Institute of Chemistry, University of Opole, Oleska 48, PL-45 052 Opole, Poland*

In random systems consisting of grains with size distributions the transport properties are difficult to explore by network models. However, the concentration dependence of effective conductivity and its critical properties can be considered within coarsened lattice model proposed that takes into account information from experimentally known size histograms. For certain classes of size distributions the specific local arrangements of grains can induce either symmetrical or unsymmetrical critical behaviour at two threshold concentrations. Using histogram related parameters the non-monotonic behaviour of the conductor–insulator and conductor–superconductor threshold is demonstrated.

## 1. Introduction

Recently, on a square lattice the model granular system has been investigated by Monte Carlo simulation [1, 2], where critical percolation coverage is shown to oscillate as a function of the size of grains. Although the authors considered only the case with monosized grains the impact of their finite size on the critical properties was clearly demonstrated in comparison to the threshold concentration for the site process (point-like grains) of a square lattice. The interesting question can be addressed to a contribution (which is due to the variety in grain sizes) to the critical properties of two-phase random granular system. The effective medium approximation (EMA) may be chosen as the simplest technique embodying the essential physics of the problem. Using the modified network extension [3] of EMA we focus on a two-dimensional lattice model. It should be mentioned here that the modified contact-conductance-distribution model [4] assuming that the site and bond problems are equivalent within the framework of EMA, demonstrates how the particle size ratio $r/R$ (large insulator particles of radius $R$ and smaller conducting particles of radius $r$) affects the value of critical volume fraction as well as the value of conductivity. Nevertheless, the lack of a lattice type model describing from different point of view the effective conductivity properties referred to a given grain size ratio but also to a grain size distribution (GSD) seems to us somewhat unsatisfactory. The purpose of the paper is to present such a model that makes use of information related to a size histogram.

## 2. Coarsened Lattice Model

Consider first a random system of 'black' and 'white' equally sized grains of high $\sigma^H$ and low $\sigma^L$ conductivity using the notation $\{1\} : \{1\}$ for GSD in this simple case. This system is represented on a square lattice in which a single grain can 'occupy' only the centre of a unit bond. Let $p$ be the unit bond fraction of phase H in a system. By definition, the only possible model bonds at this scale are the pure ones. To describe on average the influence of local arrangements of bigger H and L-grains on the effective conductivity $\sigma^*(p)$ it seems quite natural to investigate the system at larger scale. Therefore, the *coarsened* square lattice with new bonds of length $l = 2$ is chosen. To each of the bonds a model-cell is assigned with reference to the basic lattice, see Fig. 1a. Each of the model-cell consists of $l^2$ allowable 'positions' for grains of type '1'. Let us consider a more general GSD for H-grains of three sizes and monosized L-grains with the corresponding area ratios $\{4 : 2 : 1\} : \{1\}$. The grains of type '2' or '4' having two (four) times larger areas in comparison with the grain of type '1', occupy two (four) nearest centres of the model cell unit bonds, see Figs. 1b, c and d. If every 'position' in $i$-th model cell is occupied by a H-(L-phase), i.e. the local unit bond fraction of H-phase $\gamma_i = 1 (= 0)$, we have a pure H-(L-bond). Otherwise, a mixed-bond of the model occurs corresponding to $0 < \gamma_i < 1$.

---

[1]) e-mail: piaser@uni.opole.pl

In our model the simplifying assumption is proposed: all the bond conductivities depend only on the local unit bond fraction $\gamma_i$. Obviously, for different GSDs the various

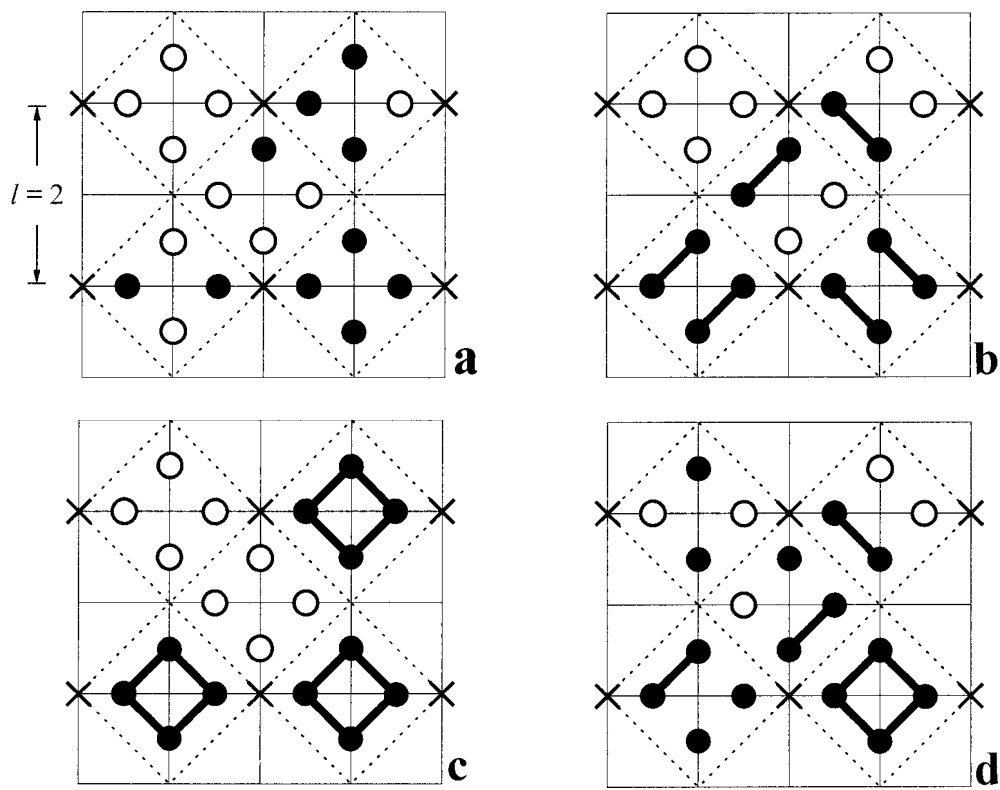

Fig. 1. Examples of the model bonds (×——×) and cells (marked by dashed lines) for different GSDs: a) $\{1\} : \{1\}$, b) $\{2\} : \{1\}$, c) $\{4\} : \{1\}$ and d) $\{4 : 2 : 1\} : \{1\}$. The L-(H-grains) of type '1' are indicated by open (filled) circles. Other H-grains of type '2' ('4') are shown as linked two (four) filled circles

local arrangements with the same $\gamma_i$ are characterized (distinguished) by their frequencies of appearance. So, at a given scale the dependence of $\sigma^*(p)$ on GSD can be described indirectly in this way.

Thus, bearing in mind the above assumption the general distribution of the bond conductivities $\sigma^H(\gamma_i = 1)$, $\sigma_i(0 < \gamma_i < 1)$ and $\sigma^L(\gamma_i = 0)$ we shall need to deal with is of the form

$$P(\sigma) = F_H(p)\,\delta(\sigma - \sigma^H) + \sum_{i \in \{\text{mixed-bonds}\}} F_i(p)\,\delta(\sigma - \sigma_i(\gamma_i))$$
$$+ F_L(p)\,\delta(\sigma - \sigma^L) \qquad (1)$$

with $F_H(p)$, $F_i(p)$ and $F_L(p)$ fractions of H-, mixed and L-bonds. According to the well-known EMA conditions for a 2D single-bond approximation [3]

$$\int P(\sigma)\,(\sigma - \sigma^*)/(\sigma + \sigma^*)\,d\sigma = 0, \qquad (2)$$

after substitution of Eq. (1) the equation for $\sigma^*(p)$ can be written as

$$F_H(p)\,\frac{\sigma^H - \sigma^*(p)}{\sigma^H + \sigma^*(p)} + \sum_{i \in \{\text{mixed-bonds}\}} F_i(p)\,\frac{\sigma_i(\gamma_i) - \sigma^*(p)}{\sigma_i(\gamma_i) + \sigma^*(p)} + F_L(p)\,\frac{\sigma^L - \sigma^*(p)}{\sigma^L + \sigma^*(p)} = 0. \qquad (3)$$

To proceed further, the mixed-bond conductivities $\sigma_i(\gamma_i)$ as well as the detailed bond fractions for a given GSD are needed. According to the previously mentioned assumption, the mixed-bond conductivities $\sigma_i(\gamma_i)$ can be calculated within the EMA again, i.e. from a corresponding quadratic expression $\sigma_i^2 + \sigma_i(\sigma^H - \sigma^L)(1 - 2\gamma_i) - \sigma^H\sigma^L = 0$. In this way we shall use in our model

$$\sigma_i = \left[-(\sigma^H - \sigma^L)(1 - 2\gamma_i) + \sqrt{(\sigma^H - \sigma^L)^2(1 - 2\gamma_i)^2 + 4\sigma^H\sigma^L}\right]\!/2. \qquad (4)$$

We can now start with introducing auxiliary functions $p_4(p)$, $p_2(p)$ and $p_1(p)$ related to the probability of occurrence of four, two the nearest 'linked' centres and one centre of model cell unit bonds occupied by a H-grain of type '4', '2' and '1', respectively. Similarly, the auxiliary function $q_1(p)$ is related to the probability of occurrence of one unit bond centre occupied by a L-grain of type '1'. Now, some correlations increasing along with $p$ caused by geometric constraints are expected between the four auxiliary functions. Considering all the allowable local configurations and the total concentration $p$ we obtain two basic model equations

$$p_4 + 2p_2^2 + 4p_2p_1^2 + p_1^4 + 8p_2p_1q_1 + 4p_1^3q_1 + 4p_2q_1^2 + 6p_1^2q_1^2 + 4p_1q_1^3 + q_1^4 = 1, \qquad (5a)$$

$$p_4 + 2p_2^2 + 4p_2p_1^2 + p_1^4 + 6p_2p_1q_1 + 3p_1^3q_1 + 2p_2q_1^2 + 3p_1^2q_1^2 + p_1q_1^3 = p. \qquad (5b)$$

The particular terms of Eq. (5a) can be identified with the model bond fractions

$$F_H(p) \equiv p_4(p) + 2p_2^2(p) + 4p_2(p)\,p_1^2(p) + p_1^4(p),$$
$$F_{0.75}(p) \equiv 8p_2(p)\,p_1(p)\,q_1(p) + 4p_1^3(p)\,q_1(p),$$
$$F_{0.50}(p) \equiv 4p_2(p)\,q_1^2(p) + 6p_1^2(p)\,q_1^2(p),$$
$$F_{0.25}(p) \equiv 4p_1(p)\,q_1^3(p), \qquad F_L(p) \equiv q_1^4(p), \qquad (6)$$

where $F_{0.75}(p)$, $F_{0.50}(p)$ and $F_{0.25}(p)$ denote the mixed-bond fractions with $\gamma_i = 0.75$, 0.50 and 0.25.

To investigate the model properties more information about the {4 : 2 : 1} : {1} GSD is necessary. For the present we wish to use a size histogram as the starting point to reduce the general set of Eqs. (5a) and (5b) to the set including only two of the four auxiliary functions. Suppose that there is a connection between the ratio of the appropriate areas for two sets of grains with different sizes and the ratio of the corresponding auxiliary functions. Then a size histogram can be always used to extract the two independent relations between $p_4$, $p_2$ and $p_1$. To check this simple computer simulations on a square lattice with 4900 bonds with attributed model cells were performed for the chosen bond concentrations. The probability of selecting a H-(L-grain) was proportional to the initial area of grains of a given type. This rule was good enough to overpass a blockage problem and not so restrictive for randomness as, for example, the sequential drawing. Further, we use a simple geometrical restriction for every model cell: the summary area of H- and L-grains was equal to the area of the cell. Then the average bond fractions were evaluated and used to find the corresponding model bond values by fitting the model parameters. For example, in case of only two different sizes of H-grains nearly linear relationships were found between the above-mentioned initial area ratios used in computer simulation and the proper model parameters $A$, $B$, $C$, $D$, $E$ and $F$ used below. Generally, one can always use our model parameters to fit all the results with those obtained from the computer simulation approach mentioned above. We shall use this much less time consuming way to investigate the critical properties of the model. At this stage we consider such simple GSDs from which the exactly solvable three limit cases {1} : {1}, {2} : {1} and {4} : {1} as well as significant insights can still be obtained.

(i) {2 : 1} : {1} GSD with $p_4 = 0$ and $p_2 = Ap_1$:

$$2A^2 p_1^2 + 4Ap_1^3 + p_1^4 + 8Ap_1^2 q_1 + 4p_1^3 q_1 + 4Ap_1 q_1^2 + 6p_1^2 q_1^2 + 4p_1 q_1^3 + q_1^4 = 1,$$
$$2A^2 p_1^2 + 4Ap_1^3 + p_1^4 + 6Ap_1^2 q_1 + 3p_1^3 q_1 + 2Ap_1 q_1^2 + 3p_1^2 q_1^2 + p_1 q_1^3 = p \qquad (7)$$

and {4 : 1} : {1} GSD with $p_2 = 0$ and $p_4 = Ep_1$:

$$Ep_1 + p_1^4 + 4p_1^3 q_1 + 6p_1^2 q_1^2 + 4p_1 q_1^3 + q_1^4 = 1.$$
$$Ep_1 + p_1^4 + 3p_1^3 q_1 + 3p_1^2 q_1^2 + p_1 q_1^3 = p. \qquad (8)$$

(ii) {2 : 1} : {1} GSD with $p_4 = 0$ and $p_1 = Bp_2$:

$$2p_2^2 + 4B^2 p_2^3 + B^4 p_2^4 + 8Bp_2^2 q_1 + 4B^3 p_2^3 q_1 + 4p_2 q_1^2 + 6B^2 p_2^2 q_1^2 + 4Bp_2 q_1^3 + q_1^4 = 1,$$
$$2p_2^2 + 4B^2 p_2^3 + B^4 p_2^4 + 6Bp_2^2 q_1 + 3B^3 p_2^3 q_1 + 2p_2 q_1^2 + 3B^2 p_2^2 q_1^2 + Bp_2 q_1^3 = p \qquad (9)$$

and {4 : 2} : {1} GSD with $p_1 = 0$ and $p_4 = Cp_2$:

$$Cp_2 + 2p_2^2 + 4p_2 q_1^2 + q_1^4 = 1,$$
$$Cp_2 + 2p_2^2 + 2p_2 q_1^2 = p. \qquad (10)$$

(iii) {4 : 1} : {1} GSD with $p_2 = 0$ and $p_1 = Fp_4$:

$$p_4 + F^4 p_4^4 + 4F^3 p_4^3 q_1 + 6F^2 p_4^2 q_1^2 + 4Fp_4 q_1^3 + q_1^4 = 1,$$
$$p_4 + F^4 p_4^4 + 3F^3 p_4^3 q_1 + 3F^2 p_4^2 q_1^2 + Fp_4 q_1^3 = p \qquad (11)$$

and $\{4:2\}:\{1\}$ GSD with $p_1 = 0$ and $p_2 = Dp_4$:

$$p_4 + 2D^2 p_4^2 + 4Dp_4 q_1^2 + q_1^4 = 1,$$
$$p_4 + 2D^2 p_4^2 + 2Dp_4 q_1^2 = p. \tag{12}$$

In order to evaluate the possible model bond fractions $F_H(p)$, $F_{0.75}(p)$, $F_{0.50}(p)$, $F_{0.25}(p)$ and $F_L(p)$ for a given model parameter, it is first necessary to solve the proper nonlinear set of equations. Then, using the general Eq. (3) the concentration dependence of effective conductivity $\sigma^*(p)$ can be calculated. This procedure can be also extended for two nonzero model parameters as well as the more general GSDs.

## 3. Results and Discussion

For comparison purposes we present first the analytical results of the model bond fractions for the limit cases mentioned above. For the case (i) after setting $A = 0$ in Eq. (7) or $E = 0$ in Eq. (8) the exactly solvable $\{1\}:\{1\}$ GSD is described by

$$\begin{cases} (p_1 + q_1)^4 = 1, \\ p_1(p_1 + q_1)^3 = p. \end{cases} \tag{13}$$

Thus, $p_1(p) = p$, $q_1(p) = 1 - p$ and we get

$$F_H(p) = p^4, \qquad F_{0.75}(p) = 4p^3(1-p), \qquad F_{0.50}(p) = 6p^2(1-p)^2,$$
$$F_{0.25}(p) = 4p(1-p)^3, \qquad F_L(p) = (1-p)^4. \tag{14}$$

For the case (ii) after setting $B = 0$ in Eq. (9) or $C = 0$ in Eq. (10) the exactly solvable $\{2\}:\{1\}$ GSD is modelled by

$$\begin{cases} 2p_2^2 + 4p_2 q_1^2 + q_1^4 = 1, \\ 2p_2^2 + 2p_2 q_1^2 = p. \end{cases} \tag{15}$$

Thus, $p_2(p) = (\sqrt{2}/2)(\Delta - (1-p))^{1/2}$ and $q_1(p) = (\Delta - p)^{1/4}$, where $\Delta \equiv (1 - 2p(1-p))^{1/2}$ and we obtain

$$F_H(p) = \Delta - (1-p), \qquad F_{0.50}(p) = 2(1-\Delta), \qquad F_L(p) = \Delta - p, \tag{16}$$

where the only with $\gamma_i = 0.50$ of mixed-bonds appears.

For the case (iii) after setting $F = 0$ in Eq. (11) or $D = 0$ in Eq. (12) the exactly solvable $\{4\}:\{1\}$ GSD is given by

$$\begin{cases} p_4 + q_1^4 = 1, \\ p_4 = p. \end{cases} \tag{17}$$

Thus, $p_4(p) = p$, $q_1(p) = (1-p)^{1/4}$ and we have

$$F_H(p) = p, \qquad F_L(p) = 1 - p, \tag{18}$$

where no mixed-bonds appear. For all the above limit cases the behaviour of the $\log \sigma^*(p)$ curve is shown in Fig. 2a. The interesting feature is the symmetrical double-threshold behaviour of $\log \sigma^*(p)$ for the limit cases $\{1\}:\{1\}$ and $\{2\}:\{1\}$ whereas for the case $\{4\}:\{1\}$ the standard threshold concentration $p_c = 1/2$ appears.

Although in our model both the conductivities are finite, i.e. $0 < \sigma^H/\sigma^L < \infty$, the appearance of mixed-bonds with a conductivity completely determined by $\sigma^H$ and $\sigma^L$ is the forcing factor to have a second threshold by the system. Of course, in the limit $\sigma^H/\sigma^L \to \infty$ only one of the precolation thresholds is observable (see the interesting remarks on the symmetrical two-conductivity thresholds of continuous medium model of the effective dielectric constant in 2D clustered mean-field approximation given by Sheng and Kohn [5]). The symmetrical two thresholds were also discussed in a three-component percolation model with nonconducting, normally conducting and highly conducting bonds for dispersed ionic conductors in [6].

A different approach for solid electrolytes was considered in [7] using continuum percolation analogue of the random resistor model which analytically describes the dependence of one of the percolation thresholds (conductor–superconductor) on the insulating particle size as well as the overall conductivity behaviour.

For the limit cases the critical behaviour can be formally obtained from Eq. (3). If conductor–superconductor (conductor–insulator) transition is considered the value of the corresponding percolation threshold $p_c^I(p_c^{II})$ is calculated by first letting $\sigma^H \to \infty (\sigma^L \to 0)$ and then setting $\sigma^*/\sigma^L \to \infty (\sigma^*/\sigma^H \to 0)$. Considering for general

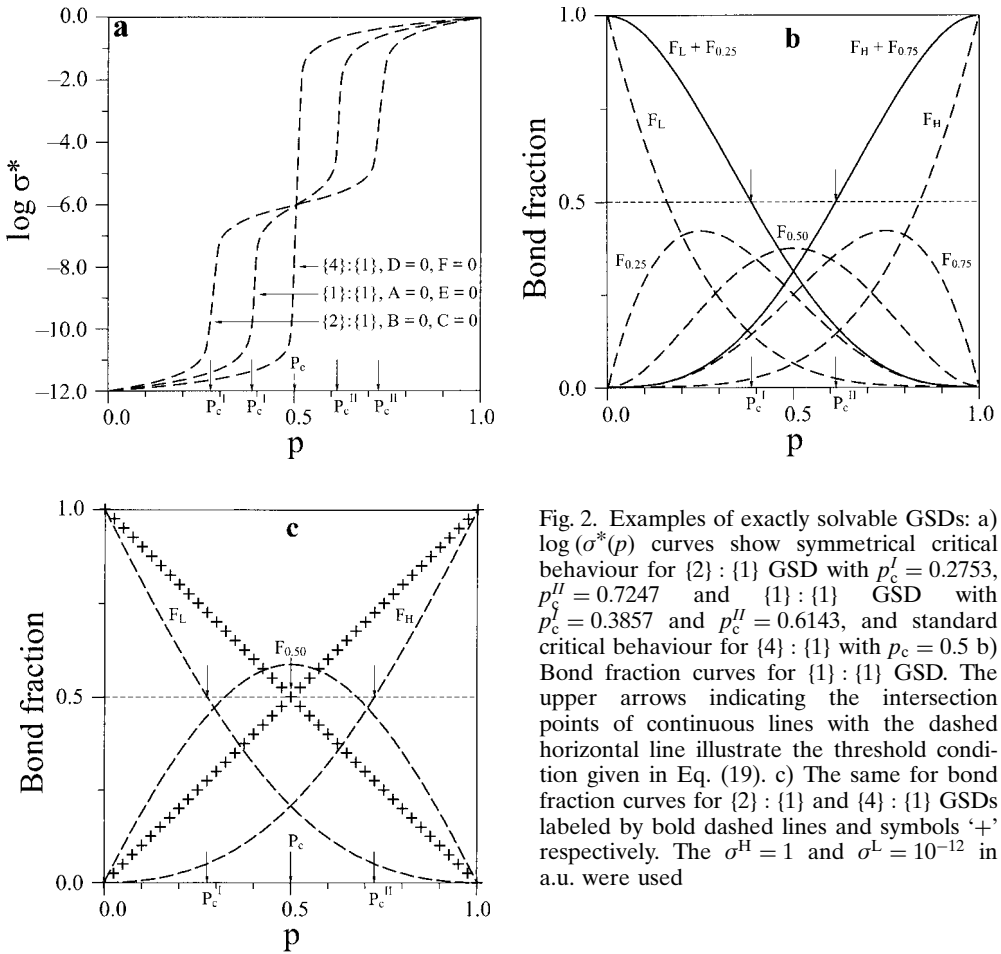

Fig. 2. Examples of exactly solvable GSDs: a) $\log(\sigma^*(p))$ curves show symmetrical critical behaviour for $\{2\} : \{1\}$ GSD with $p_c^I = 0.2753$, $p_c^{II} = 0.7247$ and $\{1\} : \{1\}$ GSD with $p_c^I = 0.3857$ and $p_c^{II} = 0.6143$, and standard critical behaviour for $\{4\} : \{1\}$ with $p_c = 0.5$ b) Bond fraction curves for $\{1\} : \{1\}$ GSD. The upper arrows indicating the intersection points of continuous lines with the dashed horizontal line illustrate the threshold condition given in Eq. (19). c) The same for bond fraction curves for $\{2\} : \{1\}$ and $\{4\} : \{1\}$ GSDs labeled by bold dashed lines and symbols '+' respectively. The $\sigma^H = 1$ and $\sigma^L = 10^{-12}$ in a.u. were used

purposes the case $\{4:2:1\}:\{1\}$ with the possible three mixed-bond conductivities $\sigma_{0.75}$, $\sigma_{0.50}$ and $\sigma_{0.25}$ and carrying out the limits the following threshold conditions are obtained:

$$F_L(p_c^I) + F_{0.25}(p_c^I) = 1/2,$$
$$F_H(p_c^{II}) + F_{0.75}(p_c^{II}) = 1/2. \qquad (19)$$

These pleasing for further discussion formulas are illustrated for the limit case $\{1\}:\{1\}$ in Fig. 2b, where the upper (lower) arrows indicate the relevant intersections (projections). With use of Eqs (14) and (19) the critical concentrations $p_c^I$ and $p_c^{II}$ can be calculated as follows:

$$6(1-p_c^I)^4 - 8(1-p_c^I)^3 + 1 = 0 \Rightarrow p_c^I = 0.3857,$$
$$6(p_c^{II})^4 - 8(p_c^{II})^3 + 1 = 0 \Rightarrow p_c^{II} = 0.6143. \qquad (20)$$

For the next limit case $\{2\}:\{1\}$ see Fig. 2c, and similarly, from Eqs (15) and (19) we get the exact values

$$4(1-p_c^I)^2 - 12(1-p_c^I) + 3 = 0 \Rightarrow p_c^I = (3-\sqrt{6})/2 = 0.2753,$$
$$4(p_c^{II})^2 - 12 p_c^{II} + 3 = 0 \Rightarrow p_c^{II} = (\sqrt{6}-1)/2 = 0.7247. \qquad (21)$$

Note that the symmetrical values of the above thresholds can be connected with the symmetry properties of bond fractions: $F_H(p) = F_L(1-p)$, $F_{0.75}(p) = F_{0.25}(1-p)$ and $F_{0.50}(p) = F_{0.50}(1-p)$. The symmetry properties are understood for the case $\{1\}:\{1\}$ with monosized grains because of the topological equivalence of the two phases. In turn, for the case $\{2\}:\{1\}$ the model system is forced to behave like for $\{2\}:\{2\}$ GSD (L-grains can occur as group of two (four) grains of type '1' so $q_1^2(q_1^4)$ plays the role of $q_2(q_2^2)$. Together with the assumption about the local dependency of bond conductivities it implies that the two phases also in this case are topologically equivalent.

The last limit case $\{4\}:\{1\}$ is shown in Fig. 2c and from Eqs. (18) and (19) the standard critical behaviour is confirmed

$$p_c^I \equiv p_c^{II} = 0.5. \qquad (22)$$

In this case the model system behaves like for $\{4\}:\{4\}$ GSD (L-grains can occur only as groups of four grains of type '1' so $q_1^4$ plays the role of $q_4$). Therefore, such 'degenerated' behaviour corresponds to the Bruggeman model with a standard binary distribution $P(\sigma) = p\,\delta(\sigma - \sigma^H) + (1-p)\,\delta(\sigma - \sigma^L)$ for the only possible bond conductivities $\sigma^h$ and $\sigma^L$.

Let us now investigate the $p_c^I$ and $p_c^{II}$ sensitivity to each of the model parameters. All that is required is solving each of the coupled Eqs. (7) to (12) and employing again the threshold conditions given by Eq. (19). The calculated curves for the pairs of parameters: $A$ for $\{1\}:\{1\} \to \{2:1\}:\{1\} \to \{2\}:\{1\}$ and $E$ for $\{1\}:\{1\} \to \{4:1\}:\{1\} \to \{4\}:\{1\}$, $B$ for $\{2\}:\{1\} \to \{2:1\}:\{1\} \to \{1\}:\{1\}$ and $C$ for $\{2\}:\{1\} \to \{4:2\}:\{1\} \to \{4\}:\{1\}$, $F$ for $\{4\}:\{1\} \to \{4:1\}:\{1\} \to \{1\}:\{1\}$ and $D$ for $\{4\}:\{1\} \to \{4:2\}:\{1\} \to \{2\}:\{1\}$ are shown in Figs. 3a, b and c. Obviously, for zero and asymptotic values of parameters symmetrical two thresholds appear. For the rest, the unsymmetrical behaviour dominates because the symmetry properties of bond fractions are lost. This observation can be connected to some deviations from randomness existing in the model system due to geometrical constraints on local ordering. Surprisingly, also non-monotonic threshold behaviour is easily seen in Figs. 3a and c. Note that the non-monotonic changes are present even as we start from the standard case $\{4\} : \{1\}$ with parameter $F$. Despite a single-parameter approach applied to particular cases the simple model system demonstrates quite interesting features. A fuller understanding of such unusual behaviour would be desirable for further extended investigations.

In general, the possible mixed-bond fractions have no more the symmetry properties needed to hold longer the relation $p_c^I = 1 - p_c^{II}$. For example, one can easily observe in Fig. 4 that two unsymmetrical percolation thresholds in $\log \sigma^*(p)$ appear for more general $\{4 : 2 : 1\} : \{1\}$ GSD described by the two model parameters $B = 1.06$ and $C = 3.21$ while for $B = C = 0$ again the symmetrical ones appear. The non-zero values of the two parameters correspond to the histogram area ratios used in computer simulation $S_1/S_2 = 0.5$ and $S_4/S_2 = 2.0$, where the subscripts refer to the types of grains. As we can see, at this stage the inclusion of real experimental data concerning the size distribution into the model needs some preparatory computer simulation work. Finally, we remark that the model proposed can be directly used to granular composites with two diverse GSDs, one for H-grains and the second for L-grains.

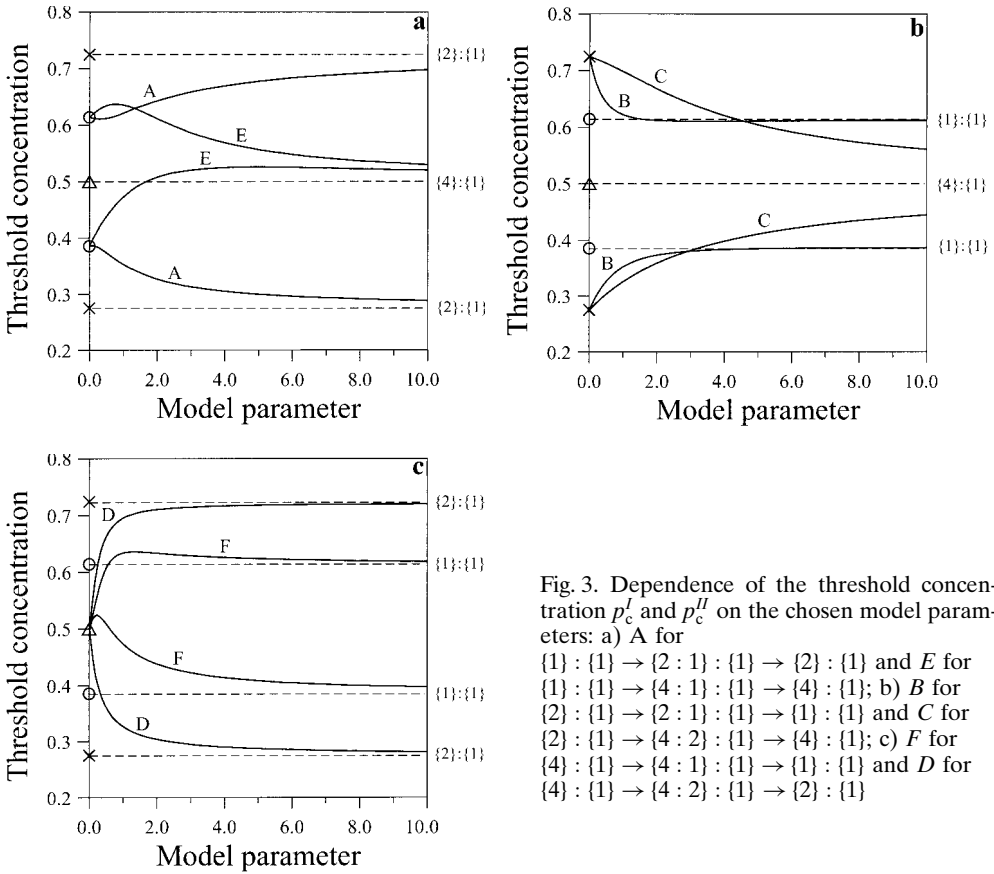

Fig. 3. Dependence of the threshold concentration $p_c^I$ and $p_c^{II}$ on the chosen model parameters: a) $A$ for
$\{1\} : \{1\} \to \{2 : 1\} : \{1\} \to \{2\} : \{1\}$ and $E$ for
$\{1\} : \{1\} \to \{4 : 1\} : \{1\} \to \{4\} : \{1\}$; b) $B$ for
$\{2\} : \{1\} \to \{2 : 1\} : \{1\} \to \{1\} : \{1\}$ and $C$ for
$\{2\} : \{1\} \to \{4 : 2\} : \{1\} \to \{4\} : \{1\}$; c) $F$ for
$\{4\} : \{1\} \to \{4 : 1\} : \{1\} \to \{1\} : \{1\}$ and $D$ for
$\{4\} : \{1\} \to \{4 : 2\} : \{1\} \to \{2\} : \{1\}$

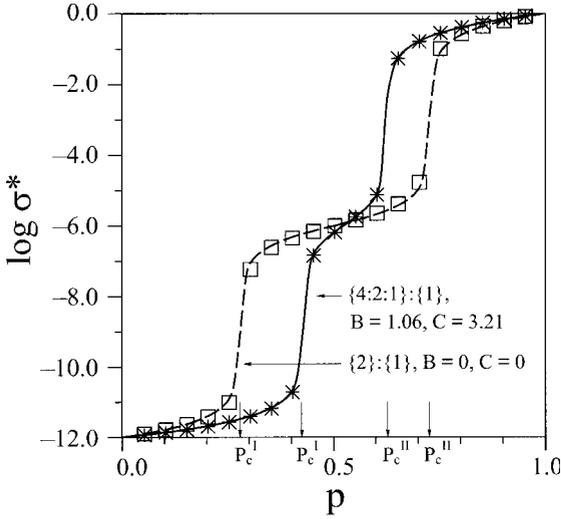

Fig. 4. Comparison of two $\log(\sigma^*(p))$ curves showing the symmetrical critical behaviour for $\{2\}:\{1\}$ GSD with $p_c^I = 0.2753$ and $p_c^{II} = 0.7247$, and the unsymmetrical one for $\{4:2:1\}:\{1\}$ GSD with $p_c^I = 0.4220$ and $p_c^{II} = 0.6250$. The symbols correspond to the values obtained by computer simulation for the histogram area ratios $S_1/S_2 = 0.5$ and $S_4/S_2 = 2.0$, where the subscripts refer to the types of grains. The $\sigma^H = 1$ and $\sigma^L = 10^{-12}$ in a.u. were used

## 4. Conclusion

In conclusion, we have presented a coarsened lattice model to describe within the effective medium approximation the influence of assumed (or known from experiment) grain size distribution on the effective conductivity and its critical properties. To take into acount the relevant features of a size histogram the appropriate one-parameter relations must be specified. This simple model demonstrates how geometrical constraints on local arrangements of grains can induce either symmetrical or unsymmetrical critical behaviour at more than one threshold. Also non-monotonic changes in threshold forced by a model parameter closely related to a size histogram have been shown.

*Acknowledgement* I would like to thank one of the referees for having drawn my attention on ref. [7].